%Paper: hep-th/9307102
%From: Palev Tchavpar <palev@ictp.trieste.it>
%Date: Thu, 15 Jul 93 18:01:00 MET DST

\baselineskip 18pt
\font\title=cmr8
\noindent
{\bf Wigner quantum oscillators}

\vskip 32pt
\noindent
T. D. Palev $^{a,b,}$\footnote*{Permanent address: Institute for Nuclear
Research
and Nuclear Energy, Boul. Tsarigradsko Chausse 72,
1784 Sofia, Bulgaria; E-mail
palev@bgearn.bitnet} and N. I. Stoilova $^{b,}$*

\noindent
$^a$ International Centre for Theoretical Physics, 34100 Trieste, Italy

\noindent
$^b$ Applied Mathematics and Computer Science, University of Ghent,
B-9000 Gent, Belgium

\vskip 32pt
We present three groups of noncanonical quantum oscillators. The position
and the momentum operators of each of the groups generate  basic Lie
superalgebras, namely $sl(1/3)$, $osp(1/6)$ and $osp(3/2)$.
The $sl(1/3)$-oscillators have  finite energy spectrum and finite-dimensions.
The $osp(1/6)$-oscillators are related to the para-Bose statistictics.
The internal angular momentum $s$ of the $osp(3/2)$-oscillators takes no
more than  three (half)integer values. In a particular representation $s=1/2$.

\vfill \eject
%\vskip 48pt

In 1950 Wigner published a paper entitled: "Do the equations of
motion determine the quantum mechanical commutation relations ?" [1].
The question to answer was whether for a (one dimensional) quantum
system with a Hamiltonian

$$H={p^2\over 2m} +V(q) $$

\noindent
one can derive the canonical commutation relations (CCR's)

$$[p,q]=-i\hbar, \eqno(1)$$

\noindent assuming that the Hamiltonian equations

$$\dot q={p\over m}, \quad \dot p=-{\partial V \over \partial q} \eqno(2)$$

\noindent
and the  Heisenberg equations (in the corresponding
picture)

$$\dot q=-{i\over \hbar}[q,H], \quad \dot p=-{i\over \hbar}[p,H]
                                                     \eqno(3)$$

\noindent
hold. The point of Wigner was that (2) and
(3) have a more immediate physical significance than (1).
The inverse is known to be true [2]: from (1) and (2) [resp.(1) and (3)]
one derives (3) [resp. (2)]. Therefore the question actually
was whether one can generalize the concept of a quantum system in
a logically consistent way. Considering as an example an
one-dimensional harmonic oscillator ($m=\hbar=1$),
$H={1 \over 2}(q^2+p^2)$, Wigner has shown that such a
generalization is possible indeed and in fact he found a family of
noncanonical solutions, labeled with an arbitrary nonnegative number
$E_0$, the energy of the ground state. In terms of the operators

$$a^+={1 \over \sqrt2}(q-ip), \quad a^-={1 \over \sqrt2}(q+ip)
                                                              \eqno(4)  $$

\noindent the result of Wigner can be stated as follows: the
Hamiltonian equations (2) are identical with the Heisenberg
equation (3) for all (representations of the) operators $a^\pm$,
which satisfy the relations [3]

$$[ \{a^\xi,a^\eta \},a^\epsilon]=(\epsilon -\xi)a^\eta
+ (\epsilon - \eta)a^\xi. \eqno(5)$$

\noindent Here and throughout $\xi, \eta, \epsilon =\pm$ or $\pm 1$;
$[x,y]=xy-yx, \; \{x,y \}=xy+yx.$ The case $E_0=1/2$ corresponds
to the canonical case, i.e., only for this value of $E_0$ $a^\pm$ are
ordinary Bose operators, $[a^-,a^+]=1$.

Although the paper of Wigner has attracted some immediate attention [4],
most of the investigations, following it, remained in the frame of
the one-dimensional case [5] (see also [3] for other references). Certainly,
one can generalize immediately the above ideas to any $n$-dimensional
oscillator, and in particular to a 3-dimensional oscillator with a
Hamiltonian

$$H={(p_1)^2+(p_2)^2+(p_3)^2 \over 2m}+
    {m \omega^2 \over 2}[(r_1)^2+(r_2)^2+(r_3)^2], \eqno(6)  $$

\noindent
assuming simply that the coordinates and momenta corresponding to  different
degrees of freedom commute with each other, $[(p_i,r_i),(p_j,r_j)]=0$
for $i \not= j$. There exist however also other, nontrivial
generalizations.  One such 3-dimensional oscillator with quite
unconventional properties was studied by one of
us (T. D. P.) in Ref. 6 (see also below). In the present note we shall
give an example of another noncanonical 3-dimensional Wigner oscillator,
which has an interesting physical property: the spin of the oscillator is
1/2. The oscillators  considered in [6,7] and the one we are
going to study here are particular cases of what we call Wigner
quantum oscillators (and, more generaly, Wigner quantum system).
The oscillator is said to be a Wigner quantum
oscillator if the following conditions are fulfilled.
\smallskip
\settabs \+  [11] & I. Patera, T. D. Palev, Theoretical
   interpretation of the experiments on the elastic \cr

\+ {\bf 1}$^0$ & The state space $W$ is a Hilbert space. The physical
                 observables are Hermitian operators in $W$.\cr
\smallskip
\+ {\bf 2}$^0$ & The Hamiltonian equations and the Heisenberg equations
                 are identical (as operator equations) in $W$.\cr
\smallskip
\+ {\bf 3}$^0$ & The internal angular momentum (the spin) of the oscillator
                 {\bf s}=$(s_1, s_2, s_3)$ is a liner function of the \cr
\+             & position operators {\bf r}=$(r_1, r_2, r_3)$ and the
                 momentum {\bf p}=$(p_1, p_2, p_3)$, so that {\bf s},
				 {\bf r} and {\bf p} transform \cr
\+			   & as vectors: $[s_j,c_k]=i \sum_{l=1}^3 \varepsilon_{jkl}c_l,
                 \quad c_k=s_k,r_k,p_k,\; i,j,k=1,2,3 $.\cr
\smallskip
\+ {\bf 4}$^0$ & The spectrum of $H$ is bounded from bellow.\cr
\smallskip
The underlying mathematical structure of the  oscillators, which we
consider, is one and the same. It is related to the representation theory of
some basic Lie superalgebras [8]. As we shall see, this is also the case
for the canonical 3-dimensional oscillator. In order to outline
the link with the Lie superalgebras (see also [6]), introduce in place
of the unknown {\bf p} and {\bf r} new unknown operators

$$a_k^\pm=\sqrt{m \omega \over 2 \hbar} r_k \mp
{i \over \sqrt {2m \omega \hbar}}p_k, \quad k=1,2,3. \eqno(7)  $$

\noindent
In terms of $a_k^\pm$, which we call creation and annihilation
operators (CAO's), the Hamiltonian (6) reads:

$$H={1 \over 2}\omega \hbar \sum_{k=1}^3 \{a_k^+, a_k^- \}.\eqno(8)  $$

\noindent
The condition {\bf 2}$^0$ yields ($k=1,2,3$):

$$\sum_{i=1}^3 [ \{a_i^+,a_i^- \},a_k^\pm]=\pm 2a_k^\pm.
                                            \eqno(9)  $$

\noindent
The equations (9) are a unique consequence from the Hamiltonian equations

$$\dot{\bf p}=-m \omega^2 {\bf r}, \quad \dot{\bf r}={{\bf
p}\over m} \eqno(10) $$

\noindent and the Heisenberg equations

$$\dot {\bf p}=-{i\over \hbar}[{\bf p},H], \quad
\dot {\bf r}=-{i\over \hbar}[{\bf r},H], \eqno(11)  $$

\noindent independently of the properties of the unknown CAO's
$a_k^\pm$. They are equal time relations, the time dependence
being $a_k^\varepsilon(t)=e^{i \varepsilon \omega t}a_k^\varepsilon(0)$,
$\varepsilon =\pm $. Hence eqs. (9) hold, if they are fulfilled at,
say, $t=0$.

In order to be slightly more general, let us denote by $F(n)$
the associative algebra with unity, generators $a_1^\pm,\ldots,a_n^\pm$
and relations (9). Then  any representation of $F(3)$ is a candidate for
a Wigner oscillator, or, more precisely, the CAO's of any Wigner
oscillator give a representation of $F(3)$. In such a case the
representation space of $F(3)$ (=the corresponding $F(3)$-module) is a
state space of the oscillator. For definiteness we call the  algebra
$F(n)$ an ($n$-dimensional) free oscillator
algebra. Thus, as a first step, one has to find the representations of
$F(3)$ and then select those of them, for which also conditions
{\bf 1}$^0$, {\bf 3}$^0$ and {\bf 4}$^0$ hold. It turns out this
is not an easy problem and, in fact, it is unsolved so far.
Here we list three classes of solutions.

\vskip 6mm
{\bf Class 1 solutions: osp(1/6) oscillators.}

\smallskip
Let $F_1(n)$ be the (free unital) associative superalgebra with
odd generators $a_1^\pm,\ldots,a_n^\pm$ and relations

$$[ \{a_i^\xi,a_j^\eta \},a_k^\epsilon]=\delta_{ik}(\epsilon -\xi)a_j^\eta
+ \delta_{jk}(\epsilon - \eta)a_i^\xi, \quad
i,j,k=1,\ldots,n, \; \xi, \eta, \epsilon =\pm \;{\rm or}\; \pm 1.\eqno(12)$$

\noindent
The operators (12) satisfy eqs. (9) and therefore $F_1(n)$ is a
factor algebra of $F(n)$. Consequently any representation of $F_1(n)$ is
a representation of $F(n)$. Observe that the solutions of Wigner
belong to this class ($n=1$). The canonical solution, namely the
one with CAO's being Bose operators is also from this class. It
is easily verified that the operators (12) are para-Bose (pB) operators
[9]. Their main algebraic property stems from the observation
that the subspace

$$B_1={\rm lin.env.}\{a_i^\xi, \{a_j^\eta,a_k^\varepsilon\} \vert
i,j,k=1,\ldots,n \; \xi, \eta, \epsilon =\pm \}
\subset F_1(n) \eqno(13) $$

\noindent
is a Lie superalgebra [10] with odd generators the pB operators. This
algebra is isomorphic to one of the basic Lie superalgebras in
the classification of Kac [8], namely to the orthosymplectic LS
$osp(1/2n) \equiv B(0/n)$, whereas $F_1(n)$ is its universal enveloping
algebra  [11]. As a result the representation
theory of any $n$ pairs of pB-operators is completely equivalent to the
representation theory of the LS $osp(1/2n)$. It is another question that
for physical reasons one has to select a subclass of
representations, which in the case $n=3$ should satisfy the conditions
{\bf 1}$^0$-{\bf 4}$^0$. Unfortunately not much is known about the
representations of $osp(1/6)$ and, more generally, about $osp(1/2n)$
(apart from the full classification of the finite-dimensional
modules [8]). The only technique
to construct new representations from the Fock representation was
developed by Green [9] through the Green ansatz [12]. It leads
however to reducible representations and is realized in tensor
products of Fock spaces. The representation with
statistics of order $p$ corresponds to the irreducible
representation of $osp(1/2n)$, containing the highest weight
vector (which is the vacuum) in the tensor product of $p$ copies of
Fock spaces (considered as $osp(1/2n)$-modules). There exists
however no effective methods to  extract this representation from
the reducible tensor product representation. This may be the reason
why the para-Bose oscillator of dimension higher than one was not
considered so far. The important for us conclusion  is that there
exist  solutions of the free oscillator algebra $F(n)$ with
operators, which generate the basic Lie superalgebras $osp(1/2n)$,
namely a LS from the class {\bf B} in the classification of
Cartan-Kac [8]. This naturally leads to the idea to try to find
solutions of eqs.
(9) with  representations of other LS's
from the same class {\bf B} or from the other classes  of
basic Lie superalgebras.

\vskip 6mm
{\bf Class 2 solutions: sl(1/3) oscillators [6].}

\smallskip
Let $F_2(n)$ be the associative superalgebra with generators
$a_1^\pm,\ldots,a_n^\pm$ and relations

$$\vcenter{\openup3\jot \halign{$#$ \hfil  \cr
[\{a_i^+,a_j^-\},a_k^+]=\delta_{jk}a_i^+ -\delta_{ij}a_k^+, \cr
[\{a_i^+,a_j^-\},a_k^-]=-\delta_{ik}a_j^- +\delta_{ij}a_k^-, \cr
\{a_i^+,a_j^+\}=\{a_i^-,a_j^-\}=0. \cr
}} \eqno(14)$$

\noindent
These operators also satisfy eqs.(10) and therefore $F_2(n)$ is another
factor algebra of the free oscillator algebra $F(n)$. The subspace

$$A={\rm lin.env.}\{a_i^\pm, \{a_j^+,a_k^- \} \vert
i,j,k=1,\ldots,n  \}
\subset F_2(n) \eqno(15) $$

\noindent
is a Lie superalgebra with odd generators $a_1^\pm,\ldots,a_n^\pm$,
which is isomorphic to the Lie superalgebra $sl(1/n) \equiv A(0,n-1)$
from the class {\bf A} of the basic Lie superalgebras. $F_2(n)$ is
its universal enveloping algebra. Hence any representation of
$sl(1/n)$ gives a solution of eqs. (9). The condition {\bf 1}$^0$
restricts the class of representations to only the finite-dimensional
representations of $sl(1/3)$, which are explicitly known [13].
The internal angular momentum (condition
{\bf 3}$^0$) is $s_i=-i\sum_{k,l=1}^3 \varepsilon_{ikl} \{a_k^+,a_l^- \}$.
A class of state spaces, labeled with any nonnegative integer $p$
was studied in [6]. The corresponding oscillator, one can call it
$sl(1/3)$-oscillator, is very unconventional. We mention some of its
properties.  The spectrum of the Hamiltonian is finite; it has no
more than 4 different eigenvalues. The square distance operator
${\bf r}^2=(r_1)^2+(r_2)^2+(r_3)^2$ is an integral of motion. Its maximal
eigenvalue is $(r_{max})^2={3p \hbar \over 2m \omega}$. Therefore the
oscillator is confined in the space. It resembles in this respect a
wavelet (see [14] and the references therein).
The spin of the oscillator is either 0 or 1. Finally, the coordinates
$r_1, r_2, r_3$ do not commute with each other, so that the position of
the oscillating particle cannot be localized. The particle is smeared
with a certain probability along a sphere with a fixed radius.

\vskip 6mm
{\bf Class 3 solutions: osp(3/2) oscillators.}

\smallskip
Another new class of solutions of the compatibility equations (9),
i.e. of condition {\bf 2}$^0$, is given with the set of all
possible representations of  operators $a_1^\pm,a_1^\pm,a_3^\pm$,
which satisfy the following relations ($\epsilon =\pm \;{\rm or}
\pm 1, \; i,j,k=1,2,3$):

$$\vcenter{\openup3\jot \halign{$#$ \hfil  \cr
[\{a_i^+,a_j^-\},a_k^\epsilon]={2 \over 3}\delta_{ik}a_j^\epsilon
-{2 \over 3}\delta_{jk}a_i^\epsilon +
{2 \over 3}\delta_{ij}\epsilon a_k^\epsilon, \cr
[\{a_i^\epsilon,a_i^\epsilon \},a_k^{-\epsilon}]=
-{4 \over 3}\epsilon a_k^\epsilon, \cr
[\{a_i^\epsilon,a_i^\epsilon \},a_k^{\epsilon}]=0, \cr
\{a_i^\epsilon,a_j^\epsilon \}=0,\; i \not= j, \cr
\{a_i^+,a_j^-\}=-\{a_j^+,a_i^-\},\; i \not= j, \cr
\{a_1^+,a_1^-\}=\{a_2^+,a_2^-\}=\{a_3^+,a_3^-\}, \cr
(a_1^\epsilon)^2=(a_2^\epsilon)^2=(a_3^\epsilon)^2.\cr
}} \eqno(16)$$

We denote by $F_3(3)$  the infinite-dimensional associative superalgebra
with generators $a_1^\pm,a_1^\pm,a_3^\pm$ and relations (16). The grading
on $F_3(3)$ is induced from the requirement that the CAO's are odd
generators. Consider the subspace

$$B_3={\rm lin.env.}\{a_i^\xi, \{a_j^\eta,a_k^\varepsilon\} \vert
i,j,k=1,\ldots,n \; \xi, \eta, \epsilon =\pm \}
\subset F_3(3) \eqno(17) $$

\noindent
and turn it into a Lie superalgebra with the natural for any associative
superalgebra supercommutator, namely
$<a,b>=ab-(-1)^{\alpha \beta}ba$, where
$\alpha=deg(a)$,  $\beta=deg(b)$.
Elsewhere we shall show that $B_3$ is isomorphic to the
orthosymplectic Lie superalgebra $osp(3/2)$ and that $F_3(3)$ is
its universal enveloping algebra. Therefore we call this oscillator
an $osp(3/2)$ oscillator. The angular momentum, satisfying condition
{\bf 3}$^0$ reads:

$$s_j=-{3i \over 4}\sum_{k,l=1}^3\epsilon_{jkl}\{a_j^-,a_k^+\},
\quad j=1,2,3. \eqno(18)  $$

The $osp(3/2)$ modules (=representation spaces) for which also the
conditions  {\bf 1}$^0$, {\bf 4}$^0$ hold are infinite-dimensional.
They are labelled with all possible pairs $(p,q)$, where $p$
is an arbitrary nonnegative half-integer, and $q$ is any negative
real number, such that $p+2q \leq 0$. All  representation spaces $W(p,q)$
(among others) have been described in [15]. The energy of the oscillator
depends only on the value of $q$ and is

$$E_n=\omega \hbar (n-2q), \quad n=0,1,2, \ldots  . \eqno(19)$$

Depending on the representation, the ground energy can be arbitrarily
close to zero (for $p=0$ and very small negative $q$), but never zero.
The spin $s$ depends mainly on $p$ and takes at most three different
values. More precisely, the spin contend within each state space $W(p,q)$
reads:
\vskip 3mm
\settabs \+  [111111] & I. Patera, T. D. Palev xxxxx & Palev  Theoretical \cr
   %sample line,  see p. 232 of the Texbook.

\+   & 1. $p=0$                 & $s=0, 1$;\cr

\+   & 2. $p=1/2,\; p+2q=0$     & $s=1/2$; \cr

\+   & 3. $p=1/2,\; p+2q<0$      & $s=1/2,\; 3/2$; \cr

\+   & 4. $p=1, \; p+2q=0$      & $s=p-1,\; p$; \cr

\+   & 5. $p\geq 1, \;p+2q<0$   & $s=p-1,\; p, \;p+1$. \cr
\vskip 3mm
\noindent
The derivation of the above results, together with the
multiplicities of the states will be given elsewhere. Here we
consider explicitly the most simple and, may be, the most
interesting representation, the one corresponding to the spin 1/2
(Case 2.). An orthonormed basis in this state space
$W(1/2,-1/4) \equiv W(1/2)$ is given with
the set of all vectors $\vert n,\; s_3) $, where $n=0,1,2,\ldots$
is labelling the energy of the state and $s_3=\pm {1 \over 2}$ is the
value of the third projection of the spin.

The transformations of the basis states under the action of the CAO's
reads:

$$\vcenter{\openup3\jot \halign{$#$ \hfil & \hskip 15pt $#$ \hfil
 \cr
 a_1^-\vert n,\; s_3)={2 \over \sqrt3}(-1)^ns_3 {\sqrt n}
 \vert n-1,\; -s_3), &
  a_1^+\vert n,\; s_3)={2 \over \sqrt3}(-1)^ns_3 \sqrt{n+1}
 \vert n+1,\; -s_3), \cr
 a_2^-\vert n,\; s_3)={i \over \sqrt3}(-1)^n {\sqrt n}
 \vert n-1,\; -s_3), &
  a_2^+\vert n,\; s_3)={i \over \sqrt3}(-1)^n \sqrt{n+1}
 \vert n+1,\; -s_3), \cr
 a_3^-\vert n,\; s_3)={1 \over \sqrt3} {\sqrt n}
 \vert n-1,\; s_3), &
  a_3^+\vert n,\; s_3)={1 \over \sqrt3} \sqrt{n+1}
 \vert n+1,\; s_3). \cr
 }} \eqno(20)$$
\vskip 5mm
{}From (8) and (20) one derives $H\vert n,\; s_3)=
\omega \hbar(n+1/2)\vert n,\; s_3).$
Thus, the energy spectrum of the oscillator in this particular
representation is the same as for an one-dimensional harmonic oscillator:

$$E_n=\omega \hbar(n+1/2), \quad n=0,1,2,\ldots .\eqno(21) $$

\noindent
The eigensubspace $W_n(1/2)$ of the Hamiltonian with energy $E_n$ is
spanned on $\vert n,\; s_3), \; s_3=\pm 1/2$ and it is closed
under the action of the spin operators. It carries
a two-dimensional irreducible representation of the spin $su(2)$
algebra with generators $s_1, s_2, s_3$. The state space $W(1/2)$
is  an infinite direct sum of spin $1/2$ modules,

$$W(1/2)=\oplus_{n=0}^\infty W_n(1/2). \eqno(22) $$

Clearly this particular $osp(3/2)$ oscillator is very different from
the canonical 3-dimensional oscillator. The next table demonstrates
this. By $W_n, \; n=0,1,2,\ldots$ we denote the eigensubspace of
the Hamiltonian with energy $E_n$.

$$\vcenter{\openup3\jot \halign{# \hfil & \hskip 15pt $#$ \hfil
& \hskip 15pt $#$ \hfil  \cr
           & {\rm Canonical \; oscillator} & Osp(3/2)\; {\rm
oscillator} \cr
Energy     & E_n=\omega \hbar(n+3/2) & E_n=\omega \hbar(n+1/2) \cr
Spin content of $W_n$ & s=n, n-2, n-4,\ldots,1 \; ({\rm or}\; 0) &
s=1/2 \cr
}} \eqno(23)$$

\vskip 6mm
The purpose of the present note was to show on simple examples
that the ideas of Wigner to study more general quantum systems,
the Wigner quantum systems in our terminology, are
very rich in their origin. If, for instance, one considers a noncanonical
two particle system with internal variables, which have the
properties of an $sl(1/3)$ oscillator [6], then the two particle system
has finite space dimensions, it behaves like a system of two (nonrelativistic)
quarks, confined in the space. The $osp(3/2)$-oscillator, viewed in
the same way, gives a model of a spin $1/2$ system, which has a classical
analog: two noncanonical point particles are curling around each other and
the resulting angular momentum of the composite system is $1/2$.
Hence this is a model of a spin (among several others;
see [16] and the references therein).

One may think that the freedom in constructing such more general
quantum systems is very large. As far as the 3-dimensional
oscillator are concerned, we may say that the oscillators considered
here exhaust all Wigner oscillators, for which the position
and the momentum operators generate simple Lie superalgebras [17].
If one goes beyond the harmonic potentials, it is an open question
to find those interactions, for which (noncanonical)   Wigner
quantum systems exists.

Elsewhere we shall show that the ideas of Wigner can be extended
for any number of particles. In particular for oscillator-like
interections between the constituents one finds solutions, for which
the composite system has finite dimensions and therefore behaves
very much like a nucleus.

\vskip 24pt
{\bf Acknowledgements}

\vskip 12pt
We are grateful to Prof.~Vanden Berghe for the possibility to work at
the Department of Applied Mathematics and Computer Science, University
of Ghent. One of us (T. D. Palev) would like to thank Professor
Abdus Salam, the International Atomic Energy Agency and UNESCO
for the hospitality at the International Centre for Theoretical Physics.
It is a pleasure to thank Dr. J. Van der Jeugt for stimulating
discussions.

This work was supported by the European Community, contract No.\
ERB-CIPA-CT92-2011 (Cooperation in Science and Technology with Central
and Eastern European Countries) and Grant $\Phi$-215 of the
Bulgarian Foundation for Scientific Research.
\vskip 24pt
\noindent
{\bf References}

\vskip 12pt
\settabs \+  [11] & I. Patera, T. D. Palev, Theoretical
   interpretation of the experiments on the elastic \cr
   %sample line,  see p. 232 of the Texbook.

\noindent
\+ [1] & E. P. Wigner, Phys. Rev. 77 (1950) 711.\cr

\noindent
\+ [2] & P. Ehrenfest, Z. Phys. 4 (1927) 455.\cr

\+ [3] & Y. Ohnuki and S. Kamefuchi, Quantum field theory and
         papastatistics,\cr
\+     & (University of Tokyo press, Springer-Verlag, Berlin 1982).\cr

\noindent
\+ [4] & L. M. Yang, Phys. Rev. 84 (1951) 788. \cr

\noindent
\+ [5] & Y. Ohnuki and S. Kamefuchi, J. Math. Phys. 19 (1978) 67;
         Z. Phys. C 2 (1979) 367;\cr

\noindent
\+     & S. Okubo, Phys. Rev. D 22 (1980) 919;\cr

\noindent
\+     & N. Mukunda, E. C. G. Sudarshan, J. K. Sharma and C. L. Mehta,
         J. Math. Phys. 21 (1980) 2386;\cr

\noindent
\+     & Y. Ohnuki and S. Watanabe, J. Math. Phys. 33 (1992) 3653.\cr

\+ [6] & T. D. Palev, J. Math. Phys. 23 (1982) 1778; Czech. J.
         Phys. B 23 (1982) 680. \cr

\+ [7] & A. H. Kamupingene, T. D. Palev and S. P. Tsaneva, J.
         Math. Phys. 27 (1986) 2067. \cr

\+ [8] & V. G. Kac, Lect. Notes Math. 626 (1978) 597. \cr

\+ [9] & H. S. Green, Phys. Rev. 90 (1953) 270. \cr

\+ [10] & M. Omote, Y. Ohnuki and S. Kamefuchi, Prog. Theor. Phys.
         56 (1976) 1948. \cr

\+ [11] & A. Ch. Ganchev and T. D. Palev, J. Math. Phys. 21
         (1980) 797. \cr

\+ [12] & O. W. Greenberg and A. M. L. Messiah, Phys. Rev. B 138
          (1965) 1155. \cr

\+ [13] & T. D. Palev, J. Math. Phys. 26 (1985) 1640; 27 (1986) 1994;
          28 (1987) 272; 28 (1987) 2280; \cr
\+      & 29 (1988) 2589;\cr
\+      & T. D. Palev, Funct. Anal. Appl, 21 (1987) 245
          (English translation).\cr

\+ [14] & A. O. Barut, in: Symposium on the foundations of modern
          physics, eds. P. Lahti and \cr
\+		& P. Mitelstaedt (World
          Scientific, 1991) p.31. \cr
\+      & A. O. Barut and A. J. Bracken, Found. Phys. 22 (1992) 1267.\cr
\+      & A. O. Barut, Phys. Lett A 17 (1992) 1.\cr

\+ [15] & N. A. Ky, T. D. Palev and N. I. Stoilova, J. Math. Phys.
          33 (1992) 1841. \cr

\+ [16] & A. O. Barut, ICTP preprint IC/93/104 (1993).\cr

\+ [17] & T. D. Palev and N. I. Stoilova, in: Classical and quantum
          systems - foundations and \cr
\+      & symmetries, eds. H. D. Doebner, W. Scherer and Fr.
          Schroeck Jr. (World Scientific, 1993) p. 318. \cr

\end